# HIGH LUMINOSITY E+E- STORAGE RING COLLIDERS TO STUDY THE HIGGS BOSON

A. Blondel, U. Geneva, Switzerland;

M. Koratzinos, Geneva, Switzerland;

R. W. Assmann, A. Butterworth, P. Janot, J. M. Jimenez, C. Grojean, A. Milanese, M. Modena, J.A. Osborne, F. Zimmermann, CERN, Geneva, Switzerland;

H. Piekarz, FNAL, U.S.A.;

K. Oide, K. Yokoya, KEK Japan;

J. Ellis, King's College London and CERN, Geneva, Switzerland;

R. Aleksan, CEA/Saclay, DSM/IRFU, Gif sur Yvette, France;

M. Klute, M. Zanetti, MIT, Cambridge, Massachusetts, USA;

M. Velasco, Northwestern U., U.S.A.;

V. Telnov, Budker INP, Novosibirsk, Russia;

L. Rivkin, EPFL, Lausanne and PSI, Villigen, Switzerland;

Y. Cai, SLAC National Accelerator Laboratory, Stanford, U.S.A.

*Abstract*

A strong candidate for the Standard Model Scalar boson, H(126), has been discovered by the Large Hadron Collider (LHC) experiments. In order to study this fundamental particle with unprecedented precision, and to perform precision tests of the closure of the Standard Model, we investigate the possibilities offered by an $e^+e^-$ storage ring collider. We use a design inspired by the B-factories, taking into account the performance achieved at LEP2, and imposing a synchrotron radiation power limit of 100 MW. At the most relevant centre-of-mass energy of 240 GeV, near-constant luminosities of $10^{34}$ cm$^{-2}$s$^{-1}$ are possible in up to four collision points for a ring of 27km circumference. The achievable luminosity increases with the bending radius, and for 80km circumference, a luminosity of $5 \cdot 10^{34}$ cm$^{-2}$s$^{-1}$ in four collision points appears feasible. Beamstrahlung becomes relevant at these high luminosities, leading to a design requirement of large momentum acceptance both in the accelerating system and in the optics. The larger machine could reach the top quark threshold, would yield luminosities per interaction point of $10^{36}$ cm$^{-2}$s$^{-1}$ at the Z pole (91 GeV) and $2 \cdot 10^{35}$ cm$^{-2}$s$^{-1}$ at the W pair production threshold (80 GeV per beam). The energy spread is reduced in the larger ring with respect to what is was at LEP, giving confidence that beam polarization for energy calibration purposes should be available up to the W pair threshold. The capabilities in term of physics performance are outlined.

# INTRODUCTION

In a seminal paper [1], B. Richter considered the limits of $e^+e^-$ storage ring colliders, and concluded that such a machine could be achieved at up to 200 GeV centre-of-mass ($E_{CM}$) energy with a luminosity of $10^{32}$ cm$^{-2}$s$^{-1}$ – the Large Electron Positron collider) (LEP) indeed achieved just that. As an alternative to $e^+e^-$ storage rings for higher energies, linear colliders have been discussed since the seminal paper of M. Tigner in 1965 [2], and U. Amaldi's proposed design in 1976 [3] looks very much like today's designs. At the time of LEP, its top energy was considered to be a crossing point with a similar luminosity for the two concepts [4], above which the linear machine would become the only possibility due to the excessive synchrotron radiation in circular machines. Since then, linear collider studies have been optimized and offer luminosities of $10^{34}$ cm$^{-2}$s$^{-1}$ at 500 GeV. Meanwhile, however, circular machines have also made considerable progress with respect to the LEP design, with the introduction of much stronger focusing at the interaction points, multi-bunching, and, after successful demonstration of top-up injection at the B-factories PEP-II and KEKB, the possibility of adding to the collider ring an independent accelerator topping up the storage ring. We should also mention the work done in the VLLC (Very Large Lepton Collider) context, a design for a 370 GeV $e^+e^-$ collider in a 233 km circumference tunnel [5] [6].

Now that the LHC experiments have discovered a new particle that is a very good candidate for the Standard Model Higgs boson [7] [8], a very important particle that should definitely be studied in great detail, we consider it interesting to revisit the possibility offered by the well-known circular machines. Indeed, LEP's $E_{CM}$ was only a fraction lower than what would have been needed for Higgs production.

The machines that we discuss, which we call LEP3 for a ring that would fit in the LHC tunnel [9] [10], or TLEP for a ring that would fit in an 80 km tunnel, is a preliminary concept that we consider so interesting that that it deserves an in-depth study, in particular because we expect it to offer a substantially better cost to luminosity ratio than a linear collider Higgs factory. The existence of serious limitations cannot yet be excluded, but it is also possible that further optimization will make it even more interesting. As matters stand, we have been able to establish for LEP3 a set of parameters for a 240 GeV centre-of-mass ring collider operating with four bunches in each beam that would be able to deliver 100fb$^{-1}$ per year to each of up to four collision points, including the multipurpose LHC experiments CMS and ATLAS as well as up to two ILC-type detectors, with a power dissipation of 50 MW per beam. The luminosity of such a machine ($10^{34}$ cm$^{-2}$s$^{-1}$ at 240GeV) is similar to what the ILC is aiming to achieve, but the ILC only has one interaction point. The ability to deliver luminosity to multiple interaction points is, in our view, a definite advantage of the circular machine over a linear one.

Soon after LEP3 was proposed, it was suggested that a longer-term strategy could be developed: if an 80km tunnel could be made available, it would be possible to repeat the successful history of the LEP/LHC tunnel. At first a circular machine based on the LEP3 design (Triple-LEP or TLEP in the following) could be hosted, with a performance even superior to LEP3 at the ZH threshold and with the added advantage of reaching the 350 GeV $E_{CM}$ $t\bar{t}$ threshold. This would also make possible a precision top-quark mass measurement and give access to the WW fusion channel to H$\nu\nu$. On a longer time scale, the 80 km tunnel could host an 80 (100) TeV hadron collider, should 16T (20T) magnets become available, or a 40-TeV collider with LHC-type magnets, and might also be used to offer ion-ion or lepton-hadron collisions at the highest energies ever seen.

Most of the measurements of the Higgs mass, width and couplings that have been advertised to be achievable at a linear collider [11] [12] could be performed with similar or even better precision at LEP3, with the exceptions of the Htt coupling and the triple Higgs coupling (see **Error! Reference source not found.**). The HWW coupling is measured with a similar precision at LEP3 and ILC(250+350), and much better by TLEP(240) [and even further improved by 50% with a run at 350 GeV]. However, we note that the Htt coupling is readily accessible with high statistics at the LHC, either directly through the associated production process or indirectly through the gg→H production mechanism and the H→γγ decay, and the HWW coupling is accessible directly via the vector-boson fusion (VBF) process, as well as via H→WW decays and indirectly via the H→γγ decay. What remains is the triple Higgs coupling, which is both very fundamental and very difficult to determine either at LHC or at a linear collider – a precision of ~±20% has been advertised for a 1.4 TeV linear collider. It will be essential to understand better the capabilities of the LHC experiments in this domain. On the other hand, the main physics process for Higgs production at an $e^+e^-$ collider at 240 GeV, the "Higgsstrahlung" process $e^+e^- \to ZH$ is unique in offering a Z tag which can be used to search for rare, unconventional or invisible H decays, thereby offering genuine direct discovery potential for such a machine.

# THE PHYSICS CASE

The aim of the LEP3 or TLEP physics programme would be a precise characterization of the Higgs mechanism. It would consist of three (or four in the case of TLEP) phases (in whichever order) with an overall duration of 5 to 10 years (longer in the case of TLEP), namely:
(i)     A Tera-Z factory at the Z peak for one year;

(ii) A Mega-W factory at the WW production threshold, for one year;
(iii) A Higgs factory at an $E_{cm}$ of 240 GeV, for five years providing $10^5$ tagged Higgs boson decays;
(iv) In the case of TLEP, studies with collisions above the t-tbar threshold.

The physics goals and feasibility have been studied in [13], with the following principal conclusions.

As shown in **Table 1,** compiled during the Higgs Factory 2012 workshop [17]. The High Luminosity LHC (HL-LHC) is already an excellent Higgs factory, especially for the couplings to bosons, top, taus and muons. The ILC at 1 TeV can do somewhat better for b couplings, and can achieve a determination of the invisible Higgs decay width and charm couplings. LEP3 in Higgs factory mode would measure accessible Higgs couplings with a precision up to two times better than at a linear collider and TLEP has a significantly better performance than any other proposal. Note that these are all preliminary projections.

**Table 1:** Expected performance on the Higgs boson couplings from LHC and e+e- colliders, as compiled from the Higgs Factory 2012 workshop **[14]**

| Accelerator → <br><br> Physical Quantity ↓ | **LHC** <br><br> 300 fb$^{-1}$ /expt | **HL-LHC** <br><br> 3000 fb$^{-1}$ /expt | **ILC** <br><br> 250 GeV 250 fb$^{-1}$ <br><br> 5 yrs | **Full ILC** <br><br> 250+350+ 1000 GeV <br><br> 5yrs each | **CLIC** <br><br> 350 GeV (500 fb$^{-1}$) 1.4 TeV (1.5 ab$^{-1}$) <br><br> 5 yrs each | **LEP3, 4 IP** <br><br> 240 GeV 2 ab$^{-1}$ (*) <br><br> 5 yrs | **TLEP, 4 IP** <br><br> 240 GeV 10 ab$^{-1}$ 5 yrs (*) <br><br> 350 GeV 1.4 ab$^{-1}$ 5 yrs (*) |
|---|---|---|---|---|---|---|---|
| $N_H$ | $1.7 \times 10^7$ | $1.7 \times 10^8$ | $6 \times 10^4$ ZH | $10^5$ ZH $1.4 \times 10^5$ H$\nu\nu$ | $7.5 \times 10^4$ ZH $4.7 \times 10^5$ H$\nu\nu$ | $4 \times 10^5$ ZH | $2 \times 10^6$ ZH $3.5 \times 10^4$ H$\nu\nu$ |
| $m_H$ (MeV) | 100 | 50 | 35 | 35 | 100 | 26 | 7 |
| $\Delta\Gamma_H / \Gamma_H$ | — | — | 10% | 3% | ongoing | 4% | 1.3% |
| $\Delta\Gamma_{inv} / \Gamma_H$ | Indirect (30%?) | Indirect (10% ?) | 1.5% | 1.0% | ongoing | 0.35% | 0.15% |
| $\Delta g_{H\gamma\gamma} / g_{H\gamma\gamma}$ | 6.5 – 5.1% | 5.4 – 1.5% | – | 5% | ongoing | 3.4% | 1.4% |
| $\Delta g_{Hgg} / g_{Hgg}$ | 11 – 5.7% | 7.5 – 2.7% | 4.5% | 2.5% | < 3% | 2.2% | 0.7% |
| $\Delta g_{HWW} / g_{HWW}$ | 5.7 – 2.7% | 4.5 – 1.0% | 4.3% | 1% | ~1% | 1.5% | 0.25% |
| $\Delta g_{HZZ} / g_{HZZ}$ | 5.7 – 2.7% | 4.5 – 1.0% | 1.3% | 1.5% | ~1% | 0.65% | 0.2% |
| $\Delta g_{HHH} / g_{HHH}$ | — | < 30% (2 expts) | — | ~30% | ~22% (~11% at 3 TeV) | — | — |
| $\Delta g_{H\mu\mu} / g_{H\mu\mu}$ | < 30% | < 10% | — | — | 10% | 14% | 7% |
| $\Delta g_{H\tau\tau} / g_{H\tau\tau}$ | 8.5 – 5.1% | 5.4 – 2.0% | 3.5% | 2.5% | ≤ 3% | 1.5% | 0.4% |
| $\Delta g_{Hcc} / g_{Hcc}$ | — | — | 3.7% | 2% | 2% | 2.0% | 0.65% |
| $\Delta g_{Hbb} / g_{Hbb}$ | 15 – 6.9% | 11 – 2.7% | 1.4% | 1% | 1% | 0.7% | 0.22% |
| $\Delta g_{Htt} / g_{Htt}$ | 14 – 8.7% | 8.0 – 3.9% | — | 5% | 3% | — | 30% |
| (*): the total luminosity is the sum of the integrated luminosity at four IPs | | | | | | | |

With LEP3 operating as a Tera-Z factory, the precision on all electroweak observables would improve by factors ranging from 25 to 100. A number of outstanding issues with the current LEP1 and SLC measurements would be resolved – as an example here we recall the 2 standard-deviation offset for the number of light neutrino species, currently at 2.984±0.008, and the 3.2 standard-deviation difference between the values of $\sin^2\theta_{eff}$ derived from $A_{LR}$ and $A_{FB}$(b) and a 2.4 standard-deviation difference between the measured value of $R_b$ and the Standard Model expectation [15]. It is worth noting that running at the Z pole will be challenging for the experiments, due to the high acquisition rate (50kHz of interesting events). Current acquisition rates, in CMS for instance, are ~1kHz, albeit with event sizes about 20 times larger than expected in LEP3 or TLEP; LHCb is planning an upgrade to 25kHz acquisition rate.

A unique feature of circular machines is the accuracy with which the beam energy can be determined. This is due to the availability of the resonant spin depolarization technique which can reach an instantaneous precision of better than 100 KeV on the beam energy. We envisage running with extra dedicated non-colliding bunches where polarization can build up and the energy measured continuously with the resonant depolarization technique [16], further improving the above precision.

Transverse polarization was measured and used at LEP up to 61 GeV per beam, limited by machine imperfections and energy spread [17]. The energy spread scales as $(E_{beam})^2/\sqrt{\rho}$ (where $\rho$ is the bending radius); beam polarization sufficient for energy calibration should therefore be readily available at TLEP up to 81 GeV, i.e. the WW threshold. A new machine with a better handle on the orbit should be able to increase this limit: a full 3D spin tracking simulation of the electron machine of the Large Hadron-electron collider (LHeC) project in the LHC tunnel resulted in a 20% polarization at beam energy of 65 GeV for typical machine misalignments [18]). Polarization wigglers would be mandatory for TLEP to decrease the polarization time to an operational value at the Z peak, as without them the polarization time would be nearly 150 hours.

This feature would allow measurements of the Z mass and width with precisions of 0.1 MeV/c$^2$ or better and the W mass with a precision of 1 MeV/c$^2$ or better.

Transverse beam polarization of 40% in collisions had been observed at LEP with one collision point with a beam-beam tune shift of 0.04, yielding a single bunch luminosity of $10^{30}$/cm$^2$/s [19]. This would translate for TLEP, taking into account of the smaller value of $\beta_y^*$ and the larger number of bunches, in a luminosity of around $10^{35}$/cm$^2$/s. In addition to the polarization wigglers, movable spin rotators as designed for HERA [20] would allow a program of longitudinal polarized beams at the Z peak, resulting in a measurement [21] of the beam polarization asymmetry with a precision of the order of $10^{-5}$ – or a precision on $\sin^2\theta_W^{eff}$ of the order of $10^{-6}$ – for one year of data taking.

LEP3 and TLEP offer a huge potential for precision measurements in the electroweak sector, orders of magnitude better than its pioneering predecessor, LEP, and much better than what can be achieved at the LHC or a linear collider. Statistically, the whole of the LEP programme could be repeated at LEP3 in one1 hour and in TLEP in 5 minutes [13]. More information on the comparison of LEP3, TLEP, the LHC and the ILC can be found in [14].

This brief summary of the physics case for LEP3/TLEP provides ample motivation to explore the concept of a large ring e$^+$e$^-$ collider.

## MAIN DESIGN CONSIDERATIONS

LEP3 and TLEP are conceived primarily as a Higgs factory with the ability of producing O(1,000,000) Higgs particles during the first years of operation. With the mass of the Higgs around 125GeV, an e+e- collider with centre-of-mass energy ($E_{CM}$) energy of 240 GeV is sufficient for achieving a Higgs production cross section within 95% of the maximum.

This energy is around 15% higher than the maximum $E_{CM}$ energy achieved at LEP2 (209 GeV). Synchrotron radiation (SR) losses scale as $(E_{beam})^4/\rho$. LEP3 at 240 GeV would consume a factor of two more energy per circulating electron than LEP2. TLEP with a larger radius could operate at an energy of 350 GeV with a similar energy loss per turn. We have limited the total power loss in the ring due to synchrotron radiation to 100 MW (50 MW per beam). This would roughly equate to a wall power consumption of 200 MW, assuming a 50% power efficiency of the RF (Radio frequency) system. Currently CERN has a 200 MW contract with France's electricity provider EdF. Energy consumption figures of the proposed LHeC project are also similar. Projected linear collider power consumptions are similar at the ZH threshold and rise up to three times higher for TeV centre-of-mass energies.

Limiting the total SR power dissipation for a tunnel with a given bending radius effectively defines the total beam current of the machine (to around 7.2mA for LEP3 at the ZH threshold with the present assumptions). This also implies that maximum luminosity is achieved by splitting the total current to as few bunches as possible Our study reveals that for LEP3 a luminosity of $10^{34}$ cm$^{-2}$s$^{-1}$ in several interaction points is possible with four bunches per beam, making the project worth pursuing.

In summary, the starting baseline assumptions and design considerations of LEP3 are as follows:
- Total centre-of-mass energy of 240GeV;
- Re-use of the LHC tunnel and available infrastructures;
- Re-use of the two general-purpose LHC experiments ATLAS and CMS;
- Limiting the total synchrotron radiation losses around the machine to 100MW;
- Maximizing the luminosity delivered to the experiments by designing a low-emittance collider ring with an efficient duty cycle and strong final focusing.

The TLEP design is an extrapolation of the same principles to a larger ring, requiring a new tunnel and infrastructure. As it turns out this is an easier machine to conceive, both from the point of view of accelerator physics and from the point of view of the external constraints.

# COHABITATION WITH THE CURRENT LHC

In the case of TLEP, tunnel excavation can commence during the physics programme of the LHC, depending on the available financing.

In the case of LEP3, the options are for the electron rings to be situated on top of the current LHC machine, as envisaged in the design brief of the LHC (options 1 and 2 below), or situated where the LHC magnets are currently residing, after the end of the LHC programme. More specifically, the options are:

1. Concurrent operation. Here both machines, the LHC and LEP3, could operate concurrently on a fill-by-fill basis and LEP3 would operate with the main LHC magnets filled with liquid Helium. This mode of operation is similar to the mode of operation needed for LHeC. It is the most challenging option in all respects, (not least because of the necessity to concurrently operate both accelerators cryogenics and vacuum systems, for instance), but is unnecessary for the physics goals of LEP3.
2. Alternating operation. Here both machines would be installed in the tunnel, with only one operating on a year-to-year or long-shutdown-to-long-shutdown basis. During LEP3 operation, the LHC magnets will not be filled with liquid Helium. Some compromises in the lattice design may be necessary to avoid interference with infrastructure components of the LHC, and a method of suspension of the LEP3 accelerator should be designed. This layout would be similar to the layout envisaged by the LHeC design study, from which we have borrowed many elements including the accelerator lattice design. No serious flaws to the LHeC proposal have been found (see section 7.8 in [18]). A possible advantage of this mode of operation might be that part of the installation could be made during shutdowns between LHC operation, decreasing the eventual installation period of LEP3.
3. Single operation. Here LEP3 would be installed after uninstalling the LHC, and could be followed by installation of higher-field magnets for High-Energy LHC operation. The disadvantage of this mode of operation is that one cannot revert to LHC operation after LEP3 (but has to continue with HE-LHC, if desired, instead). There are a series of advantages, however: much simpler logistics, a no-compromise LEP3 lattice, a flat main ring, etc., all of them offering increased performance and lower costs.

Option 1 is undesirable (and unnecessary) for LEP3; option 2 would certainly interfere with the HL-LHC operation; option 3 offers the best performance and maximum simplicity, although it delays installation and operation of LEP3 to after the completion of the LHC program. Of course, construction of LEP3 in another location than CERN could be performed somewhat earlier.

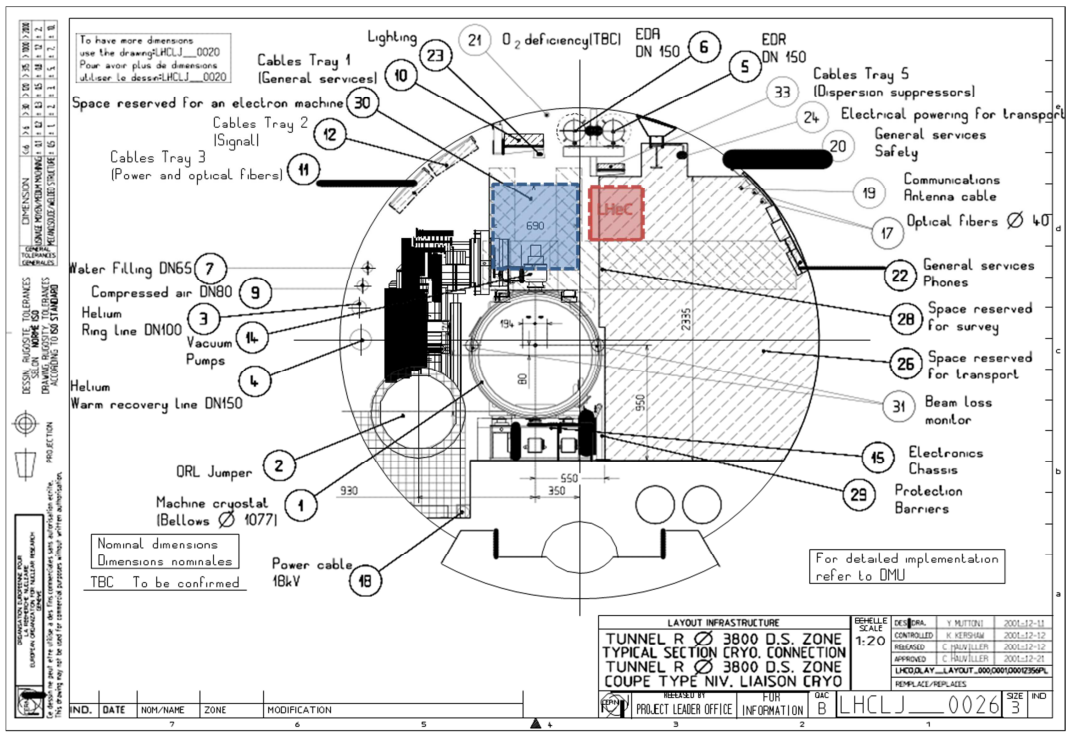

*Figure 1: A typical cross section of the LHC tunnel at a cryogenic connection representing the most stringent regular space restriction in the arcs (40 cm every approx. 50 m). In shaded blue, the area reserved for a future $e^+e^-$ colider. This provides adequate space for both the accelerator and injector rings. In red, the space envisaged for the LHeC electron ring.*

# ACCELERATOR PARAMETERS

*Figure 2* shows a schematic view of the LEP3 double ring. There is a low-emittance collider ring operating at a constant 120GeV and a second accelerator ring ramping from injection energy to 120GeV every few seconds to few minutes and "topping up" the collider ring. **Table 2** compares parameters for LEP3 and TLEP with those of LEP2 and the LHeC ring design.

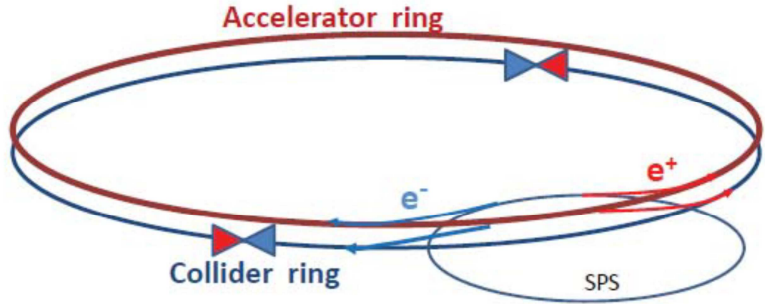

*Figure 2: Sketch of the LEP3 double ring [9]: a first ring accelerates electrons and positrons up to the operating energy (120 GeV) and injects them at few-seconds to few-minutes intervals into the low-emittance collider ring, which has high-luminosity interaction points.*

## LEP3 Parameters

We assume the same arc optics as for the LHeC, which provides a horizontal emittance significantly smaller than for LEP, at equal beam energy, and whose optical structure is compatible with the present LHC machine, allowing coexistence with the LHC. The downside of this choice is the dipole filling factor of the LHeC design which is a low 75%. This results in a smaller bending radius than that of LEP (and therefore in a larger SR loss).

Instead of the LHeC 702 MHz RF system we consider ILC-type RF cavities at a frequency of 1.3 GHz, since the latter are known to provide a high gradient and help to reduce the bunch length, thus enabling a smaller $\beta^*_y$. (but see also the discussion in the RF section below).

A key parameter is the energy loss per turn: $E_{loss}[\text{GeV}]=88.5\times10^{-6} (E_b[\text{GeV}])^4/\rho[\text{m}]$. The bending radius, $\rho$, for the LHeC is smaller than for LEP (2.6 km compared to 3.1 km), which translates into a higher energy loss than necessary. For 120 GeV beam energy the arc dipole field is 0.153 T, and a compact magnet design as in [18] can be considered.

The critical photon energy of the emitted Synchrotron radiation is 1.4 MeV. The ratio of RF voltage to energy loss per turn is increased with respect to the corresponding value at LEP in order to obtain a larger momentum acceptance. An RF gradient of 20 MV/m is considered, similar to the LHeC linac-ring design and about 2.5 times higher than for LEP. The cryogenic power increases with the square of the gradient. At 20 MV/m RF gradient, the total length of the RF sections at 120 GeV beam energy is about 20% longer than the one for LEP2 at 104.5 GeV, and the cryo power required for the collider ring is expected to be close to the current cryogenic capacity of the LHC.

The unnormalized horizontal emittance is determined by the optics and varies with the square of the beam energy. We scale it from the 60-GeV LHeC value.

The vertical emittance depends on the quality of vertical dispersion and coupling correction. The ultimate limit on the vertical emittance is set by the opening angle effect, and amounts to a negligible value, below 1 fm. We assume the vertical to horizontal emittance ratio to be similar to that for LEP. Beamstrahlung (BS) effects were estimated from analytical formulae [22] [23]. At the collision point the beams should be as flat as possible (large x/y emittance and beta ratios) to minimize energy spread and particle losses resulting from beamstrahlung [24] [25]. The bunch length of LEP3 is smaller than for LEP despite the higher beam energy, due to the smaller momentum compaction factor, the larger RF voltage, and the higher synchrotron frequency. Since LEP3 is far away from the ultimate vertical emittance limit, and since vertical emittance is so important for high luminosity, a study should be initiated to probe the physical limits that can be reached. For instance, a dynamic alignment system for the main magnets could allow the reduction of the horizontal to vertical coupling, reducing vertical emittance.

We have briefly addressed the question of the effect on vertical emittance of a possible vertical bend taking the beam from the main ring plane to the detector axis (such a bend would be needed in the case that LEP3 is installed on top of the LHC magnets). The length of a straight section where the bend could be implemented is around 1 km and the distance between the two planes around 1 m. Therefore, four vertical bends of around 2 mrad each would be needed. These bends can be provided by 40 m long dipoles, implying a bend radius of 20 kms. This is nearly a factor of 10 bigger than the bending radius of the machine in the horizontal plane. The increase on vertical emittance compared to the total horizontal emittance is proportional to the I5 integrals [26]: $\Delta\varepsilon_y/\varepsilon_x=\Delta I_{5y}/I_{5x}$. Since the $I_5$ integrals are inversely proportional to a high power of the bending radius, the effect of a vertical bump is negligible. I5 in the horizontal plane is $4\times10^{-9}$ m$^{-1}$ (LHeC optics) and $\Delta I_{5y}$ is computed to be $4\times10^{-15}$ for four vertical bends of 20 km bending radius each in an area where $\beta_y$ is around 40 m (a conservative value).

Similar to the LHeC design, the total RF wall plug power for both beams is taken to be limited to 200 MW. The wall-to-beam energy conversion efficiency is assumed to be 50%. The energy loss per turn then determines the maximum beam current. At 120 GeV beam energy it is 7.2 mA or $4\times10^{12}$ particles per beam. Additional power will be needed for the cryoplants (a total of 10-30 MW depending on the $Q_0$ value of the cavities [18]) and for the injector/accelerator rings. The total wall plug power of the LEP3 complex would then be between 200 and 300 MW (for comparison, the ILC power consumption for a centre of mass energy of 250GeV, luminosity of $0.75\times10^{34}$ and one interaction point (IP) is 125MW, so the power consumption of the two machines per Higgs produced would be similar in the case of LEP3 with two IPs).

If the total charge is distributed over 4 bunches per beam each bunch contains about $10^{12}$ electrons (or positrons), and the value of the beam-beam tune shift of ~0.09 is much less than the maximum beam-beam tune shift reached at KEKB. For comparison, LEP2 operated with a maximum beam-beam tune shift of 0.08 (at a beam energy of 94.5 GeV) without reaching the beam-beam limit, which was estimated to be 0.115/0.111 from the data taken at 98/101 GeV respectively [27].

In LEP the threshold bunch population for TMCI (Transverse Mode Coupling Instability) was about $5\times10^{11}$ at the injection energy of 22 GeV. For LEP3, at 120 GeV (with top-up injection, see below), we gain a factor 5.5 in the threshold, which cancels part of the factor $(1.3/0.35)^3 \approx 50$ increase in the magnitude of the transverse wake field of the superconducting (SC) RF cavities arising from the change in wake-field strength due to the different RF frequency. We note that only about half of the transverse kick factor in LEP came from the SC RF cavities, and that the TMCI threshold also depends (roughly linearly) on the synchrotron tune. The LEP3 synchrotron tune is about 0.35, while in LEP at injection it was below 0.15. Taking all these factors into account, LEP3 is missing a factor of 2 compared to LEP. However, the beta functions in LEP3 at the location of the RF cavities could be designed to be smaller than those in LEP (this is already true for the beta functions in the arcs), which would further push up the instability threshold. Furthermore, the LEP3 bunch length (2-3 mm) is shorter than the bunch length of LEP at LEP injection (5-9 mm). In conclusion, a bunch population of $10^{12}$ seems feasible. Alternatively, a 700MHz RF system could be envisaged giving considerably more margin (such a system is also favoured by other arguments like operation at high power throughput and the availability of an RF power source, see below.)

LEP achieved a value of $\beta^*_y$ of 4 cm. The proposed value here is 1 mm, reflecting the progress in final focus design concepts since the time of LEP (but it is still three times larger than the proposed $\beta^*_y$ of SuperKEKB). This value of $\beta^*_y$ could be realized by using new large aperture quadrupoles based on Nb$_3$Sn, as for HL-LHC (but it is not necessary to increase the field gradient by any significant amount), by a judicious choice of the free length from the IP, and possibly by a semilocal chromatic correction scheme. It is close to the value giving the maximum geometric luminosity for a bunch length of 3 mm, taking into account the hourglass effect. With a free length between the IP (Interaction Point) and the entrance face of the first quadrupole of 4 m, plus a quadrupole length of 4 m, the quadrupole field gradient should be about 17 T/m and an aperture (radius) of 5 cm would correspond to more than $20\sigma_y$, resulting in a reasonable size beam pipe around the interaction points.

At top energy in LEP2, the beam lifetime was dominated by the loss of particles in collisions [28] due to radiative Bhabha scattering with a cross section of 0.215 barn [29]. For a luminosity of $10^{34}$ cm$^{-2}$s$^{-1}$ at each of two IPs, we find a LEP3 beam lifetime of 16 minutes — LEP3 would be 'burning' the beams to produce physics very efficiently. With a LEP3 energy acceptance, $\delta_{max,RF}$, of 4%, the additional beam lifetime limit due to beamstrahlung [25] can be larger than 30 minutes, even with beams colliding at two IPs; see *Figure 3*. Assuming a beam lifetime due to beamstrahlung of 30 minutes, the overall beam lifetime would be around 10 minutes, leading to a luminosity lifetime of 5 minutes (emittance growth due to beam-beam is already included in our emittance estimates). This makes the fast top-up injection scheme indispensable if one is to have an efficient duty cycle.

In addition to the collider ring operating at constant energy, a second ring (or a recirculating linear accelerator) could be used to 'top-up' the collider; see Fig. 1. If the top-up interval is short compared with the beam lifetime this would provide an average luminosity very close to the peak luminosity. For the top-up we need to produce about $4\times10^{12}$ positrons every few minutes, or of order $2\times10^{10}$ positrons per second. For comparison, the LEP injector complex delivered positrons at a rate of order $10^{11}$ per second [30].

Concerning operation at the Z peak, again we limit the electrical power to 200 MW, with the synchrotron radiation power amounting to about half this value (with ~50% RF generation efficiency). This means that the beam current can increase $(120/45.5)^4 = 50$ times. However, the geometric emittance gets smaller at lower energy as ~energy$^2$, which would increase the luminosity as well as the beam-beam tune shift at the lower energy (by a factor 7 and 18, respectively). On the other hand, we only have a factor two margin in the tune shift. For this reason, we need to lower the charge per bunch and further increase the number of bunches. The luminosity scales as beam current times tune shift. The current increases by a factor 50. The tune shift needs to stay the same or can optimistically increase by (at the very most) a factor of 2 (from 0.08 to ~0.16). The number of bunches should therefore increase by a factor 50 times 18/2. This brings us to 920 bunches per beam for a luminosity of up to $5\times10^{35}$/cm$^2$/s at 45.5 GeV beam energy. The two limits are SR power (total current) and maximum beam-beam tune shift, which is taken to be about 0.1 per IP.

Similar scaling arguments can be used for operation at the WW threshold, resulting in about 60 bunches and $7\times10^{34}$/cm$^2$/s luminosity at a beam energy of 80 GeV.

*TLEP parameters*

A preliminary parameter list for a 350 GeV $E_{CM}$ collider (above the t-tbar threshold) in an 80-km tunnel has also been developed (see the last column in **Table 2**). The parameters were scaled from the LEP3 numbers of the 27 km ring. Also here the synchrotron radiation power is limited to 50 MW per beam. With regard to beamstrahlung the condition of Telnov [25] is just met, and according to the analytical formula a preliminary estimate of the beam lifetime from beamstrahlung may be between 15 and 30 minutes; this will need to be checked with simulations and optimized. This is less than the lifetime due to radiative Bhabha scattering of about 60 minutes (in contrast to the situation at 240 GeV $E_{CM}$ where beamstrahlung is not the limiting factor in beam lifetime). This should be long enough for efficient operation with top-up injection. The luminosity at 350 GeV $E_{CM}$ is $6.5\times10^{33}$/cm$^2$/s in each of two to four IPs. One could get more luminosity by either accepting more synchrotron radiation power, or by colliding in only one IP, or by re-optimizing the bunch parameters, for example by reducing the vertical emittance.

**Table 2**: Example parameters of LEP3 operating in Higgs factory mode and TLEP (in an 80 km tunnel) operating above the ttbar threshold and at the Z peak compared with LEP [28] [31] and the LHeC ring design [18].

| | LEP2 | LHeC | LEP3 | TLEP-ttbar mode | TLEP-Z mode |
|---|---|---|---|---|---|
| beam energy $E_b$ [GeV] | 104.5 | 60 | 120 | 175 | 45.5 |
| circumference [km] | 26.7 | 26.7 | 26.7 | 80 | 80 |
| beam current [mA] | 4 | 100 | 7.2 | 5.4 | 1180 |
| #bunches/beam | 4 | 2808 | 4 | 12 | 2625 |
| #$e-$/beam [$10^{12}$] | 2.3 | 56 | 4.0 | 9.0 | 2000 |
| horizontal emittance [nm] | 48 | 5 | 25 | 20 | 30.8 |
| vertical emittance [nm] | 0.25 | 2.5 | 0.10 | 0.1 | 0.15 |
| bending radius [km] | 3.1 | 2.6 | 2.6 | 9.0 | 9.0 |
| partition number $J_\varepsilon$ | 1.1 | 1.5 | 1.5 | 1.0 | 1.0 |
| momentum compaction $\alpha_c$ [$10^{-5}$] | 18.5 | 8.1 | 8.1 | 1.0 | 9.0 |
| SR power/beam [MW] | 11 | 44 | 50 | 50 | 50 |
| $\beta^*_x$ [m] | 1.5 | 0.18 | 0.2 | 0.2 | 0.2 |
| $\beta^*_y$ [cm] | 5 | 10 | 0.1 | 0.1 | 0.1 |
| $\sigma^*_x$ [$\mu$m] | 270 | 30 | 71 | 63 | 78 |
| $\sigma^*_y$ [$\mu$m] | 3.5 | 16 | 0.32 | 0.32 | 0.39 |
| hourglass $F_{hg}$ | 0.98 | 0.99 | 0.67 | 0.65 | 0.71 |
| $E^{SR}_{loss}$/turn [GeV] | 3.41 | 0.44 | 6.99 | 9.3 | 0.04 |
| $V_{RF}$,tot [GV] | 3.64 | 0.5 | 12.0 | 12.0 | 2.0 |
| $\delta_{max,RF}$ [%] | 0.77 | 0.66 | 4.2 | 4.9 | 4.0 |
| $\xi_x$/IP | 0.025 | N/A | 0.09 | 0.05 | 0.12 |
| $\xi_y$/IP | 0.065 | N/A | 0.08 | 0.05 | 0.12 |
| $f_s$ [kHz] | 1.6 | 0.65 | 3.91 | 0.43 | 1.29 |
| $E_{acc}$ [MV/m] | 7.5 | 11.9 | 20 | 20 | 20 |
| eff. RF length [m] | 485 | 42 | 606 | 600 | 100 |
| $f_{RF}$ [MHz] | 352 | 721 | 1300 | 700 | 700 |
| $\delta^{SR}_{rms}$ [%] | 0.22 | 0.12 | 0.23 | 0.22 | 0.06 |
| $\sigma^{SR}_{z,rms}$ [cm] | 1.61 | 0.69 | 0.23 | 0.25 | 0.19 |
| $L$/IP [$10^{32}$cm$^{-2}$s$^{-1}$] | 1.25 | N/A | 107 | 65 | 10335 |
| number of IPs | 4 | 1 | 2 | 2 | 2 |
| beam lifetime [min] | 360 | N/A | 16 | 54 | 74 |
| $\Upsilon_{BS}$ [$10^{-4}$] | 0.2 | 0.05 | 10 | 15 | 4 |
| $n_\gamma$/collision | 0.08 | 0.16 | 0.60 | 0.51 | 0.41 |
| $\Delta E^{BS}$/col. [MeV] | 0.1 | 0.02 | 33 | 61 | 3.6 |
| $\Delta E^{BS}_{rms}$/col. [MeV] | 0.3 | 0.07 | 48 | 95 | 6.2 |
| Longitudinal damping time [turns] | 34 | 200 | 26 | 19 | 1340 |
| Transverse damping time [turns] | 68 | 400 | 52 | 38 | 2280 |

## BEAMSTRAHLUNG

Beamstrahlung is an effect coined in [32] as "synchrotron radiation from a particle being deflected by the collective electromagnetic field of the opposing bunch". The effect of beamstrahlung is to limit the lifetime of the beams, as colliding leptons lose energy that takes them out of the momentum acceptance of the machine. Beamstrahlung in the circular machines is much less severe that in the linear colliders, and in particular it does not affect very significantly the collision energy or energy spread. However it introduces beam losses and must be mitigated, fault of which it can have a significant impact on the machine performance.

The effect of beamstrahlung could be very severe for beam lifetimes in case of a momentum acceptance of 2% or lower (*Figure 3* left). In this example the beam lifetime for the LEP3 chosen parameters ($\beta_x^*$ of 200mm and $100\times10^{10}$ particles per bunch) is a mere 5 minutes (compared to the 16 minute beam lifetime from Bhabha interactions). This can be mitigated by increasing the filling rate to every 30 seconds while maintaining an average luminosity of 95% of the peak luminosity. The effect can be completely eliminated, however, by a large momentum acceptance (of around 3-4% at LEP3 energies) and the tuning of vertical and horizontal emittances and beta functions at the collision points. The effect was simulated using a detailed collision simulator (Guinea-pig) that gave similar results to analytical calculations in [25]. *Figure 3* (right) shows the beam lifetime for a momentum acceptance of 4%. For TLEP this limitation is sizeable only at the highest energies.

A momentum acceptance of 4% is high compared to LEP and the LHeC (factor of six), about a factor of 2 to 3 higher than machines like superB or superKEKB, but similar to the momentum acceptance of current light sources. The momentum acceptance in the arcs is adequate (around 8%), but a careful design of the final focus system is needed that would keep the off-momentum particles over a period of the damping time (which is short). Alternative mitigations also exist: by decreasing the vertical emittance and at the same time increasing the horizontal $\beta_x^*$ we can decrease the effect of beamstrahlung while keeping the luminosity and beam-beam parameters the same.

Regarding the beamstrahlung-induced detector background, the effect is negligible compared to what an ILC detector would have to cope with. The energy spectrum of the beamstrahlung photons is much softer compared to the ILC. The number of photons, however, is higher, resulting to similar overall power dissipation. However what matters for the background to the experiments is the photon conversion to electron-positron pairs whose rate depends strongly (exponentially) on the photon momentum. In the end then, the beamstrahlung-originated e+e- pair background at a circular collider is negligible compared to ILC.

In both LEP3 and TLEP beamstrahlung will be important and therefore some effort should be directed on the one hand at understanding and simulating the phenomenon and on the other on designing a large acceptance machine.

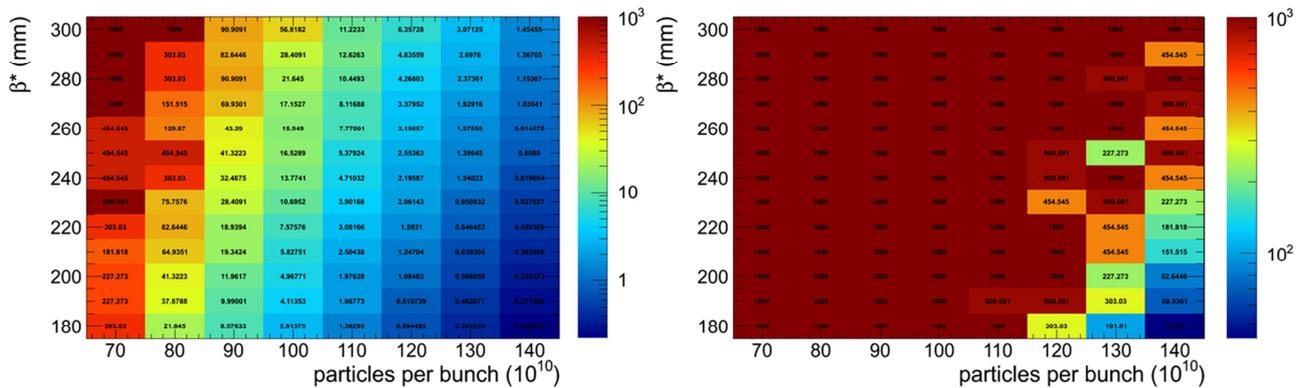

*Figure 3: Guinea-Pig simulation of the LEP3 beam lifetime due to beamstrahlung at two IPs in units of seconds (colour) versus bunch population and $\beta_x^*$ (for $\varepsilon_x$=20 nm), for a momentum acceptance of 2% (left) and 4% (right). The proposed number of particles per bunch in LEP3 is $100\times10^{10}$ and the proposed $\beta_x^*$ 200 mm. Deep red signifies lifetimes of 1000 s or more and it is a safe area for LEP3 operation.*

## DIPOLE AND QUADRUPLE MAIN MAGNETS

The best choice for the accelerator main magnets would be conventional iron-dominated electro-magnets. Even for the storage ring constantly at 120GeV there would be no advantage in using permanent magnets, considering the power consumption of the magnet system compared to the RF. Furthermore, there is an anticipated need to perform energy scans, which requires tuneable magnets.

*Per se*, the magnets of both rings would not be more challenging than the ones of the LEP era. They would need to be compatible with the emitted synchrotron radiation power and with a satisfactory field homogeneity and reproducibility. For LEP3, the one discriminating element here is that these electro-magnets would need to co-exist in the tunnel with the LHC systems installed. This sets very tight constraints, in particular on the size of these magnets. Also, the supports would need to be engineered to provide the required mechanical stability while making the overall installation feasible. These constraints do not exist for TLEP.

The LEP main bending magnets consisted of 3280 cores, each 5.75 m long [33]. The cores were made of magnetic steel in a concrete matrix, with a stacking ratio of 0.27. The gap was 100 mm, with a flux density ranging from 22 mT (20 GeV) to 110 mT (100 GeV). In the case of the proposed LEP3 machine, the 120 GeV storage ring would have a slighter lower filling ratio of the dipoles in the arcs, resulting in a flux density of 150 mT for the magnets. For the accelerator ring the corresponding magnets would have 25 mT in the gap at injection energy.

Preliminary studies have already been performed for the LHeC accelerator magnets. In the ring-ring layout, such a machine would have similar requirements for the magnets to LEP3, with in particular the same problems of co-existence with the LHC [34]. Flux densities for the LHeC dipole design range from 13 mT to 76 mT and the gap is 40 mm. A similar gap would be sufficient for LEP3. The overall dipole magnet cross-section in the LHeC design is, including the coils, around 35 cm wide and 30 cm high. Therefore, for LEP3, stacking up two units in a "double decker" configuration, would result in an overall vertical size of 60 cm. Cross-talk effects between the two magnets would need to be properly addressed, possibly with the use of trim coils. It would be challenging to further decrease the size, also in view of mechanical stability requirements. The LHeC magnet length is 5.35m. Longer magnets might be preferable for LEP3.

Regarding the main quadrupoles, compactness seems to be the main requirement. The LHeC ring-ring work can again be used as the baseline design [18]. The current LHeC ring-ring quadrupole design has a 30mm aperture, a 1 m magnetic length, a width and height of 30 cm, and a field gradient at 60 GeV of 8 or 10 T/m (different for QD/QF).

In summary, the LEP3 and TLEP accelerator magnets and in particular the bending units can benefit from the work done for the LHeC machine.

## RF CONSIDERATIONS

On the RF front, we can profit from of the enormous progress made in accelerating technology since the days of LEP2. ILC-developed cavities [35] provide a very sound possibility for a LEP3 design.

One-turn losses in LEP3 or TLEP at top energies with the current lattice taken from LHeC are 7 GeV (a higher filling factor may reduce this figure by 20% or so). However, due to the requirement for large momentum acceptance stemming from the effect of beamstrahlung, 12GV of acceleration should be made available for the main ring. This figure needs further studies to be established. We have assumed a reasonable 20 MV/m gradient, resulting in 606 m effective RF length. Much higher gradients have been achieved (the ILC uses 35 MV/m) but cryogenic power consumption goes up with the square of the gradient. This corresponds to 580 TESLA cavities @ 1.038 m per cavity, or to 73 cryomodules with 8 cavities per module (XFEL type). This in turn corresponds to a total length of 818 m compared to the 864 m of LEP2, i.e. an RF system comparable in size.

Regarding the efficiency of the RF system (the ratio of wall plug power to power delivered to the beam), we have assumed the conservative value of 50%. The LHeC design (section 8.3.2 of [18]) uses a 95% efficiency in the power converters, a 65% klystron efficiency, and waveguide transmission losses of 7%, leading to an overall efficiency of 55%.

The next question is the cryogenic power needed for these superconductive cavities: At maximum operating gradient, and assuming the worst-case cavity quality factor $Q_0$ for the TESLA cavities of $1.0 \times 10^{10}$ [36], the cryogenic load at 2K is 3 kW per sector, compared with the LHC installed capacity of 2.1 or 2.4 kW per sector. However, this $Q_0$ value is probably pessimistic, and it is possible that the LHC capacity would be sufficient.

Like in LEP2, distributing the RF power over four straight sections matches the cryogenic power availability and minimizes the energy sawtooth.

Regarding the RF power needed, for 100 MW of total beam power at 7.2 mA per beam with 580 cavities, a continuous power of 172 kW per cavity will be required. 1.3 GHz klystrons are currently available for pulsed operation at up to about 150 kW average power [37] [38], and further design work will be required to produce a klystron for continuous wave operation above this power level. One area where current technology is not adequate is in the power couplers, where L-band coupler designs are currently limited to continuous power levels of around 60 kW [39]. The choice of a lower RF frequency such as 700MH would allow higher klystron average power levels, and the larger physical size of the power couplers and RF distribution components leads to a design which is generally more robust, less sensitive and less challenging.

## VACUUM AND BEAM PIPE IN THE PRESENCE OF SYNCROTRON RADIATION

The vacuum engineering of the beam pipe is critical for a machine where the SR power loss is so high, about 5kW per meter in the arcs. Huge developments have been made on how to handle the SR power for the new generations of SR Facilities (Diamond, Soleil, Alba, MAX IV, etc.). However, the experience gathered at LEP2 is unique and useful for this project. Many lessons were learned – At LEP2, most of the issues were induced by fast changes of the beam orbit which resulted is unexpected high power SR losses which in turn created fast temperature rise and pressure bumps (photon stimulated desorption combined with thermal outgassing). The fast temperature gradient lead to leaks due to the differential thermal expansion inside the vacuum interconnections. Also, 50% of the SR power escaping in the tunnel did create severe problems like degradation of organic material (cable insulation) and damage to electronics due to high dose rates and the formation of ozone and nitric acid leading to corrosion problems. Similarly to what was done for LEP, the heat load will be extracted by water circulation. The power being higher, the beam-pipe cross section being smaller and the lead shielding required being thicker, careful simulations are needed to validate this technical solution.

Regarding the choice of material for the beam pipe, material other than that used for LEP could be considered. At LEP, extruded aluminium was used due to its good thermal conductivity and ability to extrude complex shapes. But this choice also presented a series of limitations: the pressurised hot water for the bake-out was limited to 150°C since the reliability of the vacuum interconnections based on aluminum flanges was a concern at higher temperatures (>150°C); this excluded the use of NEG coatings which have minimum activation temperature of 180°C; corrosion problems mean that materials and brazing fluids should be carefully selected. Stainless steel is an alternative that should be investigated offering a higher resistance to corrosion, reliable vacuum connections and thinner beampipe walls. It also has, however, worse heat and electrical conductivity as compared to aluminium, and is more expensive to machine and shape. The choice of material for the beam pipe remains an open question.

The critical energy of the synchrotron radiation grows as $(E_{beam})^3/\rho$. For LEP3, it is 1.5 times that for LEP2. The thickness of the lead shielding required to achieve a similar protection will be twice as thick: 6 mm on top and bottom and 12 mm where the SR is impacting. The situation is easier for TLEP since the critical energy is always less than for LEP2. Regarding vacuum, heavy gasses with a higher ionisation cross section are more harmful than Hydrogen: Argon is 67 times more harmful than Hydrogen. At LEP, the vacuum cleaning took 500 hours since NEG coating technology was not available. This vacuum cleaning will be shortened if NEG coatings (which also offer very low photon stimulated desorption coefficients) are used as baseline.

In conclusion, the LEP3 or TLEP beam pipe, shielding and heat extraction should be part of an integrated design, and the wealth of information gathered at LEP2 or in the B-factories should be of great help.

## FINAL FOCUSING

The requirements for LEP3 and TLEP regarding the final focusing elements are similar. Compact design and compatibility with operation a few meters from the IP and inside a particle detector are needed. The field strength needed is of the order of 20T/m. This is a modest field gradient compared to CLIC proposals (which are a factor of 20 higher). The aperture requirement is 10cm. This leaves the option of superconducting, conventional iron electromagnetic (EM), permanent magnet (PM) or hybrid (PM/EM) as the technology choice.

The superconducting option (with magnetic field strength and quality dominated by the very precise coils winding and positioning) is a well-known and mastered technology, being widely adopted for the LHC magnets and also proposed for the HL-LHC upgrade.

A conventional iron-dominated electromagnetic design it is probably not the most appropriate choice due to the weight and the dimensions of the magnets and the fact that they should be placed inside the detectors (L*~ 4 m).

The hybrid PM/EM solution could be very interesting due to:
a) the compactness and lightness of the magnets (so minimizing the solid angle subtracted to the Detector volume)
b) the limited need of service systems and connections (powering, water and/or cryogenic cooling)
c) the absence of cryostats and consequently of thermal contractions, an aspect that simplifies the precise alignment of the quadrupoles, with a direct strong impact on the achievable luminosity.

In this area there are recent R&D activities at CERN for the LINAC4 and CLIC projects. As an example, *Figure 4* (left) shows PM Quadrupole prototype built for LINAC4 project. Its inner magnet bore is 45 mm and its Nominal Gradient 16 T/m. This magnet is not remotely tunable, but it is very compact and the field quality would be appropriate for use in a collider like LEP3.

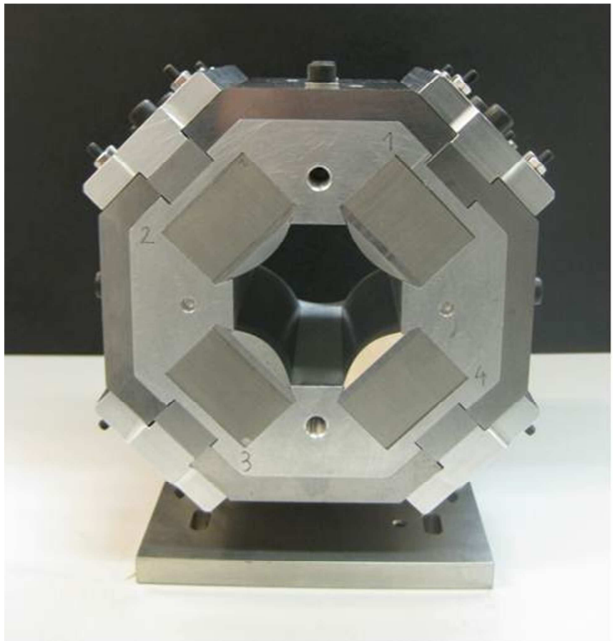 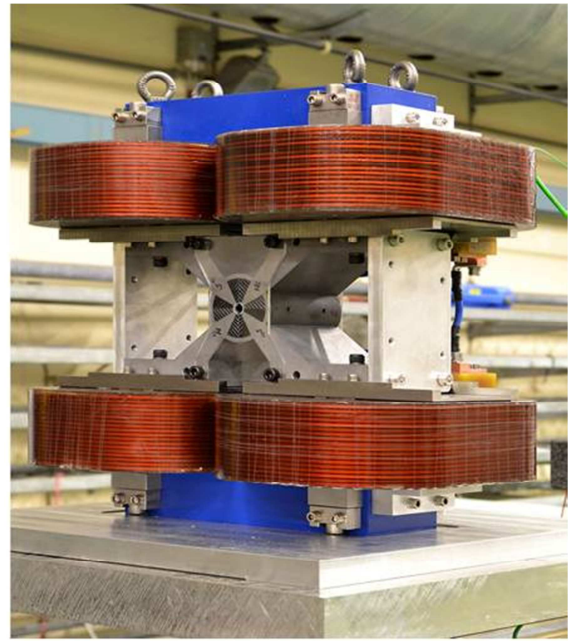

*Figure 4: LEFT: Prototype of a Permanent Magnet Quadrupole developed at CERN for the LINAC4 Project. RIGHT: Short prototype of a Tunable Hybrid Final Focus quadrupole built in the framework of CLIC R&D.*

Stronger gradients and large remote tunability are possible if one introduces in the design EM coils and with the use of high saturation materials like Permendur.

*Figure 4* (right) shows another example, a Tunable Hybrid Final Focus quadrupole developed in the framework of CLIC R&D. The inner magnet bore is 8.12 mm and the maximum gradient is 515 T/m. Tunability is about 70%. Very good field quality was achieved mainly due to the choice of producing the most critical components using the technique of Electrical Discharge Machining EDM.

If the hybrid option is favoured, a solution like the one presented in *Figure 4* (left) but with higher gradient and with the addition of tuning coils could be studied so that it satisfies the LEP3 requirements and at the same time is as small and lightweight as possible. A similar design option is investigated for a possible upgrade of the ATF2 Final Focus doublets for KEK in Japan.

## INJECTION AND FILLING

Roughly constant luminosity and beam currents will be realized through a top-up injection scheme as used at KEKB and PEP-II. Although the LEP injection apparatus has been dismantled, the tunnels linking the SPS and the LHC tunnel are available to be re-used. Therefore, a design similar to the injection scheme used for LEP can be envisaged. An injection energy of around 20 GeV would mean that there will be a factor 6-9 between injection and top energy. The LEP injector complex was able to provide $O(10^{11})$ electrons/positrons per second. LEP3 or TLEP would require $4\times10^{12}$ electrons/positrons every 16 minutes (or 1000 seconds). This is, therefore, not a very challenging requirement for the injection system.

The nominal ramp rate of LEP was 500 MeV/s [40] (although 125 MeV/s was mostly used). At the same speed, acceleration to 120 GeV would take around 3 minutes. However, the SPS when used as a LEP injector had a much shorter cycle: it accelerated $e^{\pm}$ from 3.5 to 20 GeV (later 22 GeV) in 265 ms or about 62.GeV/s. Extrapolated to LEP3 or TLEP this corresponds to an acceleration time (20 to 120 GeV) of 1.6 s. Therefore, a total cycle time of 10 s does not look very challenging. During this period of 10 s the beam intensity of a machine with a 16 min beam lifetime would have decreased by 1%. The number of electrons and positrons needed during the 10 s cycle is $4\times10^{10}$, smaller than the rate available at LEP. As done at PEP-II the sensitive parts of the detectors probably need to be gated to be blind during injection for a few tens of turns, of the order of the transverse radiation damping time.

## MODIFICATIONS TO EXPERIMENTS AND PERFORMANCE

As a case study, the use of an LHC detector on the $e^+e^-$ machine has been considered. A recent study of the CMS experiment concludes that CMS performance in LEP3 would approach those of a dedicated linear collider detector, should the vertex detector be upgraded to enable c-quark tagging [13]. ATLAS is expected to show similar performance, although no dedicated study has been performed yet.

Two integration issues merit closer investigation: the first is the presence of a final focusing quadrupole inside the detector. The first set of parameters for this quadrupole are that it needs a modest 19T/m gradient, it has a length of 4 m and needs to be placed 4m from the IP. Integration issues do not seem extremely challenging with such a device and a possible positioning of such a device inside CMS is shown in *Figure 5*.

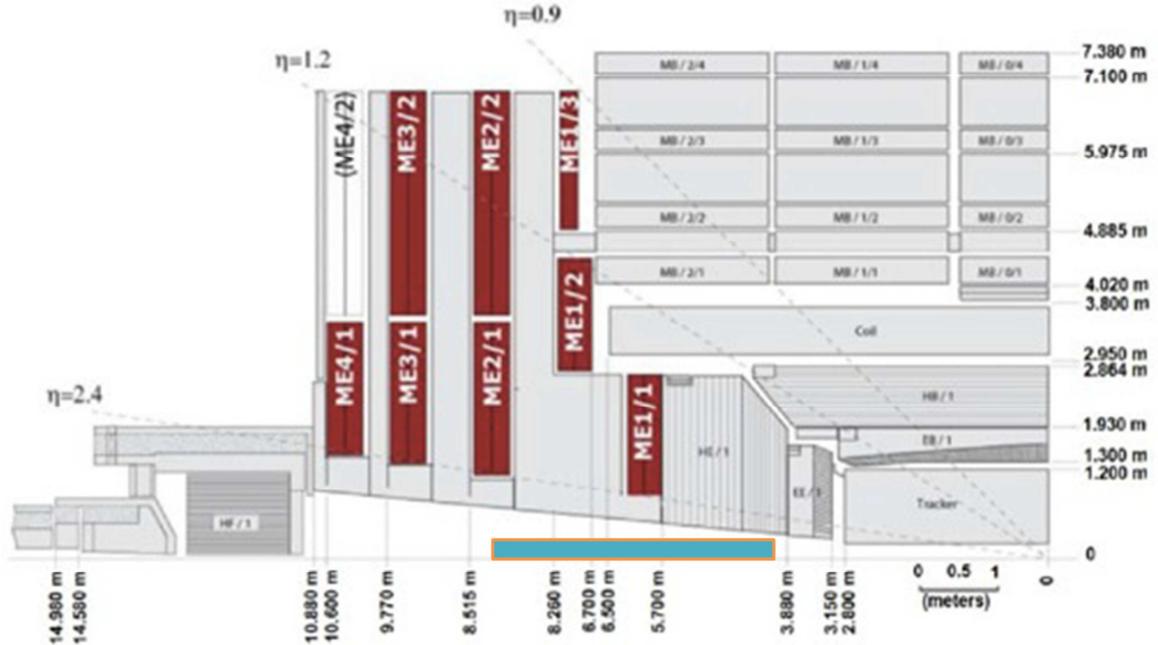

*Figure 5: Longitudinal cut of the CMS detector. The final focusing quadrupole is seen in blue 4 m from the IP. Its length is 4m, its radius is 30 cm and the radius of the beam pipe is 5cm.*

The second issue is the way that the accelerator beam will pass through or by-pass the detector. A bypass has no integration issues for the detector, but is the more expensive option. A pass-through can be integrated in two ways: the first would be using a separate beampipe that would cross the detector some distance from the main beam pipe (the distance of the two beampipes in the arcs is around 30cm). The second would be to use a common beam pipe for both beams close to the IP, with the accelerator beam coming at a small angle with respect to the horizontal or vertical planes. Both of these solutions have important integration issues that should be looked at carefully to see if any of them is feasible.

It was concluded that CMS can be an excellent $e^+e^-$ detector with a performance for Higgs physics comparable to linear collider detectors. Regarding integration issues, a more comprehensive study should take place.

## TWO RINGS VERSUS A ONE-RING DESIGN

The baseline for the LEP3 accelerator is a double ring design due to a manifold of reasons: first, a ring kept constantly at the same energy will always have superior performance to a ring that is ramped. Secondly, the average luminosity of such a machine is kept very close to the maximum luminosity, something especially important in a machine where the beam lifetime is short (16 minutes with two IPs) compared to the expected duty cycle of such an accelerator.

On the other hand, a one-ring design has the edge on cost and avoids integration problems or expensive bypasses of the accelerator beams around the experiments.

To be able to calculate exactly how much performance the second ring buys us, we need to make assumptions about the duty cycle of such an accelerator. For LEP2, physics-to-physics times were reduced to a bit less than an hour [41], ramping from injection energy to top energy taking around 13 minutes at 125MeV/s acceleration. If LEP3 could manage half this time, i.e. 30 mins, from the duty cycle alone, the average luminosity would be a factor of 0.14 of the peak luminosity (assuming 16 minute beam lifetimes or 8 minute luminosity lifetimes)

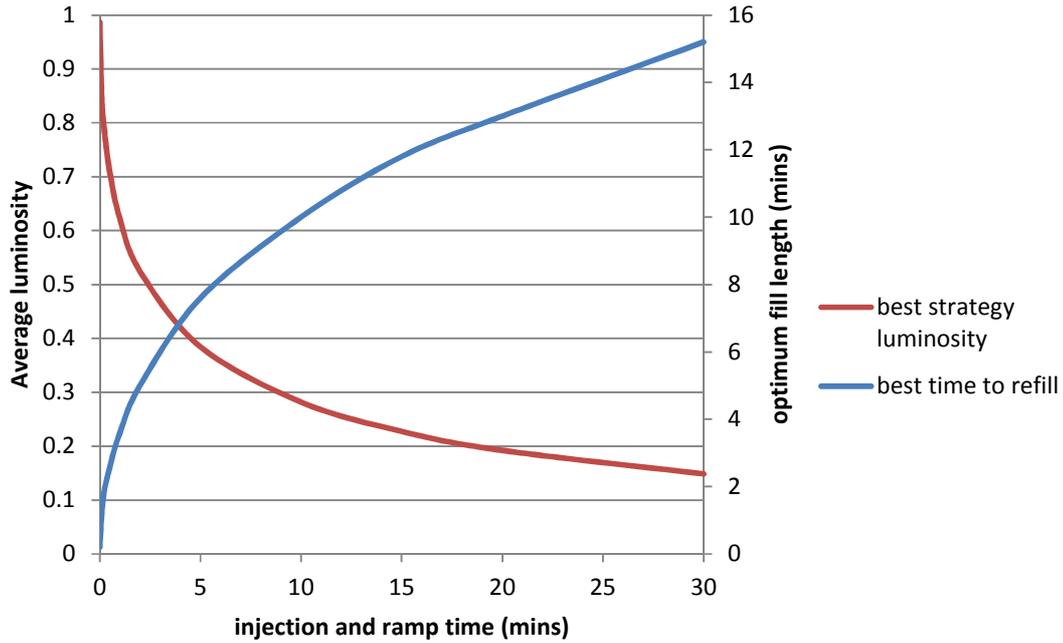

*Figure 6: A toy simulation of the effect of the duty cycle to the average luminosity (in red, the best strategy average luminosity, in blue the best strategy time in a fill to refill). Horizontal axis is the time needed for a refill, during which no collisions take place. A two-ring approach has virtually zero refill time. If the time needed to refill is 30 mins, average luminosity is decreased by a factor 7.*

Figure 6 shows the effect of different duty cycles on the average luminosity of the machine, assuming no luminosity levelling and with a beam lifetime of 16 minutes (or a luminosity lifetime of 8 minutes). If the refill time is milliseconds, then one should refill continuously, and the average luminosity is 1. Refilling every minute one loses some 7% of this theoretical maximum value. This is the case with a two-ring design. On the other hand, if filling (and all the necessary steps until physics conditions are reached, including ramping to the collision energy) takes 30 minutes, then the best strategy is to refill 15 minutes into the fill, achieving an average luminosity which is 14% of the maximum. To be able to approach to within 80% of the luminosity of the two ring design, one needs to be able to fill in 12 seconds, and needs to do so every 1.8 minutes. In conclusion, the short luminosity lifetimes expected in LEP3 or TLEP at their top energies require a two-ring design or extremely fast fill and ramp times, of the order of a few seconds. It is expected that the top-up mode would lead to more stable operation of the collider ring and, from the performance point of view, is much preferred.

## ALTERNATIVE COLLIDERS

LEP3 in the LHC tunnel is not the only possible collider to be used as a Higgs factory, if the cost of excavation of a new tunnel can be contemplated. We here mention other possible circular colliders, existing or proposed, in the Geneva region or elsewhere.

The UNK ring in Russia near Moscow still exists. It is a large bore tunnel, but its circumference is 20 km, inferior to that of LEP3.

A project similar to LEP3, called SuperTRISTAN, has recently been proposed in Japan [42]. This would be a 40 km or a 60 km tunnel in the vicinity of KEK in Japan with a luminosity similar to the one of LEP3. In the very rough cost estimate in [42] more than half of the cost of the project would be for the tunnel and the detectors.

Another possibility, which we call 'DLEP', is to build a new larger tunnel, e.g. of twice the LEP circumference, which could later be used to accommodate a High-Energy LHC with 40 TeV centre-of-mass energy. Rings with circumferences up to 50 km were considered during the LEP design in the 1970s with part of the tunnel located in the rocks of the Jura, 800-900 m under the crest [43]. Recent studies actually disfavour a 54-km ring in the Geneva region in favour of a larger 80-km or a smaller 46-km one for reasons that have to do with the morphology of the crust in the Geneva region [44]. Nevertheless our studies show that by relaxing a lot of the parameters of LEP3 (SR power per unit length, beamstahlung, beam-beam tune shift) we still end up with a machine achieving luminosities higher than $10^{34}$ cm$^2$s$^{-1}$ in each of two IPs, which outperforms a linear collider. Obviously pushing the parameters higher would result in even better performance.

The largest tunnel considered is an 80 km tunnel that we have labelled TLEP. Such a tunnel could be constructed in a number of sites around the world (CERN, Fermilab, KEK, etc.). A newly commissioned study for an 80km tunnel in the Geneva region has published its preliminary findings in [44]. An 80 km tunnel paves the way for a 350 GeV $E_{cm}$ machine above the top-quark pair-production threshold and its associated precision measurements. Such a tunnel could be used for a very-high-energy proton accelerator as well.

# REQUIRED R&D AND SYNERGIES

Storage-ring colliders represent a well-established robust technology. Nevertheless, LEP3 or TLEP are not easy machines, but must master a number of challenges. Novel features are the high momentum acceptance compared to LEP2; top-up injection, requiring a dedicated accelerator ring to sustain near-constant luminosity; low vertical $\beta^*$ (which is still 3-4 times larger than the design $\beta^*$ value for the two Super B factories); heating and stability issues for short bunches with high bunch charge; and operation in a regime of significant beamstrahlung [24] [25].

The machine parameters need to be further optimized. This is a process that has just started. A non-exhaustive list of important R&D items for LEP3 and TLEP includes:

- beam dynamics studies and optics design for the collider ring; HOM heating with large bunch currents and very small bunch lengths (0.3 cm), vertical emittance tuning, single-bunch charge limits, longitudinal effects associated with a $Q_s$ of 0.35, low beta insertion with large momentum acceptance, parameter optimization, beam-beam effects including beamstrahlung, and the top-up scheme;
- optics design and beam dynamics for the accelerator ring, and its ramping speed;
- the design and prototyping of a collider-ring dipole magnet, an accelerator-ring dipole magnet, and a low-beta quadrupole;
- 100 MW synchrotron radiation effects: damage considerations, energy consumption, irradiation effects on LHC and LEP3 equipment, associated shielding and cooling;
- Superconducting RF integration, cryogenics design and prototyping (possibly in synergy with SPL and LHeC),
- determining the optimum RF gradient as a compromise between cryo power and space, and the optimum RF frequency with regard to impedance, RF efficiency and bunch length;
- Polarization studies both for transverse and longitudinal polarization at the Z peak, plus transverse polarization at the W threshold. Study of polarization wigglers;
- engineering study of alternative new tunnels for TLEP (and HE-LHC);
- cost and performance comparison for the proposed double ring and for a single combined ring;
- design study of the LEP3/TLEP injector complex, including a positron source, and a polarized electron source;
- 3-D integration in the LHC tunnel and possible cohabitation with HL-LHC and LHeC (if the need arises).
- study of a dual use ring for LEP3 and LHeC;
- machine-detector interface, e.g. the integration of warm low-beta quadrupoles inside the ATLAS and CMS detectors, as well as a thorough detector background study;
- detector performance and upgrade studies for LEP3/TLEP, suitability of the existing LHC detectors (or the desirability of new ones) for e+e- physics – this is well under way – and additional equipment needed (low beta insertions and luminosity monitors); and
- LEP3/TLEP physics studies (also well under way) [13].

The priorities for the immediate short term would be the design of a large momentum acceptance final focus system, a custom-made lattice and a study of radiation shielding issues.

Development of arc magnets and 3-D integration can profit from synergies with the LHeC. Also part of the RF development could proceed together with similar activities for HP-SPL and LHeC. LEP3 RF cavities and RF power sources could be also used for an ERL-based LHeC (or vice versa).

# SUMMARY

The parameter list of **Table 2** allows us to draw several encouraging conclusions. It is possible to envisage an electron-positron collider in the LEP/LHC tunnel or a larger ring with reasonable parameters operating at 120 GeV per beam with a peak luminosity higher than $10^{34}$ cm$^{-2}$s$^{-1}$ in each of up to four interaction points leading to integrated luminosities of more than 100 fb$^{-1}$/yr per experiment, while keeping the total synchrotron radiation power loss below 100 MW. The beam lifetime is short, so that a good efficiency calls for a machine combining a storage ring and an independent accelerator that tops up the storage ring. While the storage ring collider technology is very mature and benefits from the progress made on LEP2, the B-factories and synchrotron light sources, a substantial effort in design and R&D was outlined. High statistics on the recently discovered H(126) boson, the Z peak and the W boson would be accumulated, allowing an exciting program of possible discovery of new physics and tests of the electroweak sector with unprecedented accuracy.